\documentclass{jfm}
\usepackage{graphicx}
\usepackage{epstopdf, epsfig}
\usepackage{soul}
\usepackage{xcolor}
\usepackage{bm}
\usepackage{amsmath}
\usepackage{url}

\newcommand{\D}{\partial}

\newcommand{\bv}[1]{\bm{#1}}
\newcommand{\vv}{\bv{v}}
\newcommand{\OO}{\bv{\Omega}}
\newcommand{\BB}{\bv{B_0}}
\newcommand{\bb}{\bv{b}}

\newcommand{\OoRo}{Ro^{-1}}
\newcommand{\opara}{\hat{\bv{x}}^\Omega_\parallel}

\newcommand{\bpara}{\hat{\bv{x}}^{B_0}_\parallel}

\shorttitle{MHD turbulence with misaligned rotation and guide field}
\shortauthor{S. J. Benavides, K. J. Burns, B. Gallet, J. Y-K. Cho and G. R. Flierl}

\title{Inverse cascade suppression and shear layer formation in MHD turbulence subject to a guide field and misaligned rotation}

\author{Santiago J. Benavides\aff{1}
  \corresp{\email{santib@mit.edu}}, Keaton J. Burns\aff{2,3}, Basile Gallet\aff{4}, \\ James Y-K. Cho\aff{3} \and
  Glenn R. Flierl\aff{1}
	}

\affiliation{
	\aff{1}Department of Earth, Atmospheric, and Planetary Sciences, Massachusetts Institute of Technology, Cambridge, MA 02139, USA
	\aff{2} Department of Mathematics, Massachusetts Institute of Technology, Cambridge, MA 02139, USA
	\aff{3} Center for Computational Astrophysics, Flatiron Institute, New York, NY 10010, USA
	\aff{4} Universit\'{e} Paris-Saclay, CNRS, CEA, Service de Physique de l’Etat Condens\'{e}, 91191 Gif-sur-Yvette, France
}

\begin{document}

\maketitle


\begin{abstract}
Astrophysical flows are often subject to both rotation and large-scale background magnetic fields. Individually, each is known to two-dimensionalize the flow in the perpendicular plane. In realistic settings, both of these effects are simultaneously present and, importantly, need not be aligned. In this work, we numerically investigate three-dimensional forced magnetohydrodynamic (MHD) turbulence subject to the competing effects of global rotation and a perpendicular background magnetic field. We focus on the case of a strong background field and find that increasing the rotation rate from zero produces significant changes in the structure of the turbulent flow. Starting with a two-dimensional inverse energy cascade at zero rotation, the flow first transitions to a forward cascade of kinetic energy, then to a shear-layer dominated regime, and finally to a second shear-layer regime where the kinetic energy flux is strongly suppressed and the energy transfer is mediated by the induced magnetic field. We show that the first two transitions occur at distinct values of the Rossby number, and the third occurs at a distinct value of the Lehnert number. The three-dimensional results are confirmed using an asymptotic two-dimensional, three-component model, which allows us to extend our results to the planetary-relevant case of an arbitrary angle between the rotation vector and guide field. More generally, our results demonstrate that, when considering the simultaneous limits of strong rotation and a strong guide field, the order in which those limits are taken matters in the misaligned case.
\end{abstract}





\section{Introduction}\label{sec:intro}
Turbulence in geophysical and astrophysical settings contains additional physical ingredients that break the isotropy of the flow, a traditional assumption in classical turbulence theory, thereby adding complexity to the system at hand  \citep{FrischBook,Davidson2013,Alexakis_Review}.
This includes, but is not limited to, rotating, electrically conducting, stratified, and large aspect ratio systems.
Asymptotic regimes are sought out to simplify the system, thus allowing previous ideas and techniques of idealized turbulence to be used. This is done by studying the limiting equations as a control parameter (rotation, aspect ratio, etc.) is taken to zero or infinity.
For example, one particular success is the quasigeostrophic approximation, which predicts horizontal motion in the presence of stratification and rapid rotation \citep{Charney1971,Vallis2017}. More generally, in rapidly rotating systems without stratification and with periodic boundary conditions, the flow becomes two-dimensional (2D), invariant along the rotation axis \citep{Smith1999,Mininni2010,Gallet2015,Buzzicotti2018}.
A similar simplification occurs in plasmas in the presence of a strong uniform background magnetic `guiding' field, reducing the dynamics to 2D magnetohydrodynamics (MHD) \citep{Montgomery1981, Nazarenko2007,Bigot2011,Alexakis2011,Sujovolsky2016}, and further to 2D hydrodynamics (HD) if the magnetic field is not forced \citep{Alexakis2011,Sujovolsky2016}. 
Both of these limits produce 2D HD turbulence, which is characterized by the presence of an \textit{inverse cascade} of energy, in which energy goes from the forcing scale towards larger scales \citep{Kraichnan1967,Boffetta2012,Alexakis_Review}. This is in contrast to the forward energy cascades found in 3D HD and MHD turbulence in the absence of a guiding field, where energy cascades to smaller scales. 
The asymptotic regimes allow one to use energy cascade arguments to help understand turbulent geophysical and astrophysical phenomena. For example, the inverse cascade in the quasigeostrophic system is thought to contribute to the formation of jets in rapidly rotating planetary atmospheres \citep{Rhines1975,Held1996,Cho1996a,Cho1996b,Arbic2004,Tobias2007,Gallet2021}. An analogous cascade mechanism is thought to be responsible for the formation of poloidal jets in tokamak plasmas in the presence of a strong background toroidal guiding magnetic field \citep{Diamond2005}.

In many geophysical and astrophysical contexts, however, it is expected that a fluid is subject to some combination of rotation, magnetic field, and stratification \citep{Cho2008,Davidson2013,Vallis2017}. Asymptotic analysis of these combined cases is more difficult, where often the order in which the limits are taken matters, and knowing which regime is observed in nature (and how the energy cascades behave) is a challenge \citep{Aurnou2015}.
%
%
%
%
%
%
Furthermore, real physical systems are not subject to infinite rotation rates or infinite background magnetic field strengths and reality often lies at intermediate values.
There is currently no existing theory for the cascade direction of such intermediate parameters, and it is only more recently through state-of-the-art simulations \citep{Smith1996,Smith1999,Celani2010,Pouquet2013,Deusebio2014,Marino2015} and lab experiments \citep{Xia2011,Campagne2014,Baker2018} that we are able to carefully investigate their turbulent dynamics. These studies looking into the cascade of conserved quantities in geophysical and astrophysical flows have revealed the presence of \textit{bidirectional cascades}\footnote{Not to be confused with \textit{dual} cascade scenarios, where the system has two conserved quadratic quantities which cascade in different directions, such as in 2D HD turbulence with the forward cascade of enstrophy and inverse cascade of energy.} at intermediate parameter values, in which a fraction of the conserved quantity input by the forcing goes to large scales whereas the rest goes to small scales \citep{Alexakis_Review,Pouquet2019}. Most of these systems seem to form bidirectional cascades at particular \textit{critical} values of the control parameters. Numerical simulations are crucial in revealing the behavior of turbulent systems in configurations and parameter values that are out of reach of asymptotic methods.

Here we investigate the turbulent dynamics of an incompressible electrically-conducting MHD fluid subject to rotation and a \textit{misaligned} uniform background magnetic field using a series of direct numerical simulations. 
Such a configuration is expected to represent the turbulent dynamics in the atmospheric interiors of gas giant planets in the transition region between the outer, neutral atmosphere and the deep, ionized one (e.g., \cite{Liu2008,Dietrich2018,Benavides2020}). There, the dynamics are characterized by rapid rotation and the presence of a strong background field generated by the dynamo in the deep interior region below. A simplified case of a dipole magnetic field present in the transition layer would suggest that the alignment between rotation and the background field would vary with latitude. The latest Juno measurements by \cite{Moore2018} show, however, that the background field around the transition region is quite `patchy', but we still expect the misalignment with rotation to be a generic feature. In these regions the electrical conductivity is expected to be quite low \citep{Liu2008,French2012,Dietrich2018}. For the sake of generality, in the following we investigate a model with rather large conductivity, before discussing how most of the results carry over to the low-conducting case in Section \ref{sec:conclusions}. To some extent, the ultimate effect of the background magnetic field is the same, resulting in anisotropic flows, and eventually the two-dimensionalization of the flow perpendicular to the field  \citep{Sommeria1982,Vorobev2005,Thess2007,Favier2010,GalletDoering2015,Baker2018}.
While our interests are at the fundamental level, with application to gas giant planets in mind, the effects of a background field and (possibly misaligned) rotation also need to be considered in the formation and dynamics of ionized protoplanetary disks in the presence of the host star's magnetosphere \citep{Fromag2005,Armitage2011,Joos2012,Simon2013,Simon2018}. Both of the astrophysical settings mentioned so far are geometrically confined, so we will not explore large domain size effects in this work (see discussion in Section \ref{sec:2D}).

More generally, given the prevalence of astrophysical systems which are both ionized and undergoing rotation, we expect our results to be general enough to apply in other contexts. Our idealized system has simplified forcing and boundary conditions compared to realistic astrophysical settings. However, its role is to uncover the dynamics of the small scales, which can eventually guide parametrizations of sub-grid scale fluxes in large-scale models of astrophysical objects.

In particular, we are interested in understanding what happens when there are two, two-dimensionalizing effects which act in different directions. What is the fate of the inverse cascade and how `fragile' is it to the variation in the secondary control parameter? 
Focusing on the case of a strong background field, we find that increasing the rotation rate from zero produces significant changes in the structure of the turbulent flow. Starting from a two-dimensional inverse cascade scenario at zero rotation, we find four distinct dynamical regimes as we increase rotation: for weak rotation rates we observe a bidirectional cascade of kinetic energy, with energy flux to large scales decreasing as rotation is increased, and negligible induced magnetic energy.
For rotation rates past some critical point, the flow transitions to a purely forward cascade of kinetic energy. 
Further increasing the rotation rate results in a shear-layer dominated regime, where nonlinearities at large scales are suppressed. 
Finally, at the largest rotation rates we investigated, we found a second shear-layer regime where the induced magnetic energy is no longer negligible, the kinetic energy flux is strongly suppressed, and the energy transfer is purely mediated by nonlinear terms which include the induced magnetic field. Using a two-dimensional, three-component asymptotic model of our system, we also show that the first three regimes are separated by sharp transitions, hinting at the existence of a bifurcation in the behavior of the turbulent flow. One is found to be similar to other previously-found transitions from a bidirectional cascade to a forward one, while the other shows subcritical behavior including a discontinuity in the order parameter and hysteresis. The transition to the magnetically active regime is beyond the scope of the reduced model, but we show that it also sharpens towards a critical value as the background magnetic field strength increases. We find more generally that, when considering the limit of strong rotation and strong magnetic field, the order in which those limits are taken matters.

In section \ref{sec:RMHDB} we introduce the system we will study: rotating MHD in the presence of a background magnetic field, referred to as $B\Omega$-MHD \citep{Menu2019}. In section \ref{sec:3D} we discuss results from three-dimensional simulations in which the background magnetic field is strong and as we vary the rotation rate in a perpendicular direction. In section \ref{sec:2D} we introduce a two-dimensional, three-component (2D3C) asymptotic model (similar to that derived in \cite{Montgomery1981}) representing the strong background magnetic field limit and including rotation, and discuss results from the simulations of that reduced system. Discussion and implications of our results are presented in section \ref{sec:conclusions}, where we extend our results to an arbitrary angle between rotation and background magnetic field, before discussing the low conductivity limit, relevant to planetary settings.

\section{Rotating MHD in the presence of a background magnetic field}\label{sec:RMHDB}
The equations for rotating magnetohydrodynamics in the presence of a uniform background magnetic field are \citep{Shebalin2006,Galtier2014}:
\begin{align}
	\frac{\D \vv}{\D t} + (\vv \cdot \nabla) \vv &= - \nabla p^*  - 2 \bv{\Omega}\times \vv + \left( \nabla \times \bb \right) \times \left(\bv{B_0}+\bb\right) + \bv{D}_v + \bv{f}, \label{eq:RMHDB1} \\
	\frac{\D \bb}{\D t} + (\vv \cdot \nabla)\bb &= (\bv{B_0} \cdot \nabla)\vv + (\bb \cdot \nabla)\vv + \bv{D}_b, \label{eq:RMHDB2} \\
    \nabla \cdot \vv = 0, &\quad \nabla \cdot \bb = 0, \label{eq:RMHDB3}
\end{align}
where $\vv = (v_x,v_y,v_z)$ is the velocity field and $\bb$ is the induced magnetic field, making up the two dynamical variables in this system. The two control parameters are $\bv{\Omega}$, the global rotation vector (with magnitude $\Omega$), and $\bv{B_0}$, the uniform background field (with magnitude $B_0$). Other definitions include the total pressure modified by rotation $p^*$, which is normalized by the constant density $\rho_0$, and the dissipation terms, $\bv{D}_v$ and $\bv{D}_b$, which could be regular viscosity and magnetic diffusion, respectively, but might also include other forms of dissipation such as drag or hypodiffusion. The exact form of these terms will be described in Section \ref{sec:3D}, when the simulations are introduced. Magnetic fields are in Alfv\'{e}n units, being normalized by $\sqrt{\rho_0 \mu_0}$, where $\mu_0$ is the magnetic permeability. Finally, $\bv{f}$ is a body force, which will be used to inject energy into the velocity field.

The inviscid and perfectly conducting system conserves the total energy,
\begin{equation}
    E = \frac{1}{2} \int\left(\vv^2+\bb^2\right) \ d^3x.
\end{equation}
However, when $\OO$ and $\BB$ are collinear, this system also conserves what's known as the \textit{parallel-helicity} \citep{Shebalin2006} or \textit{hybrid-helicity} \citep{Galtier2014,Menu2019}. The collinear system has received considerable attention -- favored over the misaligned case in part due to its extra conserved quantity and the potential relevance of its cascade for dynamo action\citep{Shebalin2006,Menu2019}. It also possesses simplified linear wave solutions which have been used to develop a weak wave turbulence theory \citep{Galtier2014,Bell2019}. Here we will \textit{not} be considering the collinear case, and so only the total energy will be conserved in our study of $B\Omega$-MHD in section \ref{sec:3D}. Although waves are certainly present in our system, our work concerns the strongly turbulent dynamics of energy cascades (present partly in the zero frequency modes of the system). See Appendix \ref{app:waves} for the dispersion relation of waves in the misaligned case.

Most studies, with rotation and background magnetic field aligned or not, have focused on how rotation and a moderate background field affect the decay of kinetic and magnetic energies in unforced simulations \citep{Lehnert1955,Favier2012,Bell2019,Baklouti2019}. \cite{Menu2019} investigated the sensitivity of the cascade of hybrid helicity for various rotation and guide field alignments in forced-dissipative simulations. We consider the effects of rotation and a misaligned background magnetic field on the two-dimensionalization of the flow and the energy cascade, including the limits of strong rotation and strong background magnetic field. 

In our study, the rotation and background magnetic field vectors are perpendicular to each other, namely, we have chosen $\OO = \Omega \bv{\hat{z}}$ and $\BB = B_0 \bv{\hat{x}}$, the extension to an arbitrary angle between $\OO$ and $\BB$ being discussed in Section \ref{sec:conclusions}. The turbulence is maintained at a statistically steady state by a forcing which inputs energy at a mean rate $I$ at a length-scale $1/k_f$ (see details in section \ref{sec:3D}). As a result, there is an emergent velocity scale $U$ defined to be $U^3 \equiv I k_f^{-1}$, that we compare to the background field as a measure of its strength, the inverse Alfv\'{e}n Mach number:
\begin{equation}
    M^{-1} \equiv \frac{B_0}{U}. \label{eq:VA}
\end{equation}
This dimensionless number can also be thought of as a measure of how the third term on the right hand side of equation (\ref{eq:RMHDB1}) (the Lorentz force) and the first term on the right hand side of equation (\ref{eq:RMHDB2}) compare to the advection terms in each respective equation, which would determine whether or not the background field affects the dominant dynamics. When $M^{-1} \gg 1$ the Lorentz force acts to constrain the velocity and induced magnetic fields so that they don't vary along the $x$-direction and most of the energy lies in the $k_x = 0$ modes, resulting in 2D MHD \citep{Montgomery1981,Nazarenko2007,Bigot2011,Alexakis2011,Sujovolsky2016}. It is important to note that while the dynamics depend on $y$, $z$, and $t$, all vector components can be non-zero in periodic domains. This is called two-dimensional, three-component (2D3C) dynamics\citep{Biferale2017}. If the induced magnetic field isn't directly forced (as is the case in our study), this results in 2D3C HD and an inverse cascade of horizontal kinetic energy \citep{Alexakis2011,Sujovolsky2016}.
All of our simulations lie in the regime of strong background magnetic field, $M^{-1} \gg 1$, making the rotation rate the main control parameter in our study. 
Does this asymptotic regime survive in the presence of rotation? 

The relative strength of rotation is measured by the inverse Rossby number:
\begin{equation}
    Ro^{-1} \equiv \frac{2 \Omega}{k_f U}. \label{eq:OoRo}
\end{equation}
This number measures the relative importance of the second term on the right hand side of equation (\ref{eq:RMHDB1}) (the Coriolis force) to the advection term, which would determine whether or not the rotation affects the dynamics. Unlike the background magnetic field, the Coriolis force only directly affects the velocity field. For non-stratified rapidly rotating hydrodynamics in the absence of any magnetic field, $\OoRo \gg 1$, the strong Coriolis force acts to constrain the flow such that it doesn't vary along the $z$-direction and most of the energy lies in the $k_z = 0$ modes \citep{Smith1999,Mininni2010,Gallet2015,Vallis2017,Buzzicotti2018}, which results in 2D3C HD where the dynamics depend only on $x$, $y$, and $t$. If the fluid is ionized and initialized with a non-zero seed magnetic field, rapid rotation doesn't necessarily result in 2D3C HD because there is no direct constraint on the induced magnetic field. Instead, if the transverse velocity component does not vanish, rapidly rotating dynamos are formed with $z$-dependent induced magnetic fields \citep{Otani1993,Smith2004,Aurnou2015,Seshasayanan2016_2d5,Seshasayanan2017,Tobias2021}.
However, since our base state is the $x$-independent 2D3C HD regime found when $M^{-1} \gg 1$, rapid rotation is expected to act to constrain the flow and prevent it from varying in the $z$-direction. Note that, in this configuration, rotation is in the plane of the 2D dynamics, not out of the plane as is often the case when it itself is the cause of the bidimensionalization. 
Since rotation is now in the plane of the two-dimensional velocities, the Coriolis force is expected to deflect horizontal velocities out of the plane, as will be discussed in section \ref{sec:2D} when we introduce a reduced model for this system following \cite{Montgomery1981}.

Our goal in this study is to investigate the effects that in-plane rotation has on the two-dimensional flow caused by a strong background magnetic field. In the next section we will describe results from direct numerical simulations of the $B\Omega$-MHD system for various rotation rates, paying particular attention to the resulting energy cascade and morphology of the flow field.


\section{Strong background field limit: 3D $B\Omega$-MHD simulations}\label{sec:3D}
Equations (\ref{eq:RMHDB1})--(\ref{eq:RMHDB3}) were solved numerically in a triply-periodic domain of side length $2 \pi L$ using the Geophysical High-Order Suite for Turbulence (GHOST) code \citep{Mininni2011}. The dissipation terms, $\bv{D}_v$ and $\bv{D}_b$, each consist of a `hyperviscosity' and a large-scale dissipation term called `hypoviscosity'. The hyperviscosity replaces the regular viscous and magnetic diffusion terms with a Laplacian of a higher order, in our case $\nabla^2 \rightarrow -\nabla^4$. This higher order allows for the possibility of forcing at smaller length-scales while still properly resolving the smallest scales at moderate resolutions. As long as the order of the Laplacian is not very large, hyperviscosity has been shown to have no significant effect on the turbulent properties of 3D turbulence, and we expect the same to be the case for our work \citep{Agrawal2020}. The hypoviscosity, which would appear as $\nu_- \nabla^{-2} \vv$ on the right hand side of equation (\ref{eq:RMHDB1}) and as $\eta_- \nabla^{-2}\bb$ on the right hand side of equation (\ref{eq:RMHDB2}), acts as a large-scale dissipation term. 
The resulting expressions for $\bv{D}_v$ and $\bv{D}_b$ are,
\begin{align*}
    \bv{D}_v &= -\nu \nabla^4 \vv + \nu_- \nabla^{-2}\vv, \\
    \bv{D}_b &= -\eta \nabla^4 \bb + \eta_- \nabla^{-2}\bb,
\end{align*}
where $\nu$ is the kinematic `hyper'-viscosity, $\eta = (\mu_0 \sigma)^{-1}$ is the magnetic `hyper'-diffusivity, $\sigma$ is the electrical conductivity, $\nu_-$ is the `hypo'-viscosity, and $\eta_-$ is the magnetic `hypo'-diffusivity. 
Should an inverse cascade of a conserved quantity occur, this term ensures that no condensate forms, which would otherwise affect the cascades and inertial ranges \citep{Chertkov2007,Xia2008,Gallet2013,Seshasayanan2018,Alexakis_Review,VanKan2019}. This is done by choosing the coefficients $\nu_-$ and $\eta_-$ such that the kinetic and magnetic energy at the largest scales is smaller than that of the next largest scales. The modified GHOST code which includes these alternative dissipative terms can be found in a public Github repository \citep{Benavides_Code}. It is a standard parallel pseudo-spectral code with a fourth-order Runge–Kutta scheme for time integration and a two-thirds dealiasing rule. The numerical model is nondimensionalized by $L$ and the forcing amplitude $f_0$, so that the wavenumbers $\bv{k}$ correspond to mode numbers of the domain and the forcing amplitude is one. The three-dimensional forcing $\bv{f}$ is isotropic and constant in time, comprising of a summation of cosines with wavenumbers between $8 < |\bv{k}| < 10$, making $k_f$ = 9, and random phases. The forcing wavenumber range is chosen in an attempt to properly resolve both an inverse cascade and a forward cascade. $I \equiv \langle \bv{f} \cdot \vv \rangle$ is the space- and time-averaged energy injection rate, where $\langle \cdot \rangle$ represents a space- and time-average. We do not force the induced magnetic field.

\begin{figure}
	\centering{
		\includegraphics[width=0.48\textwidth]{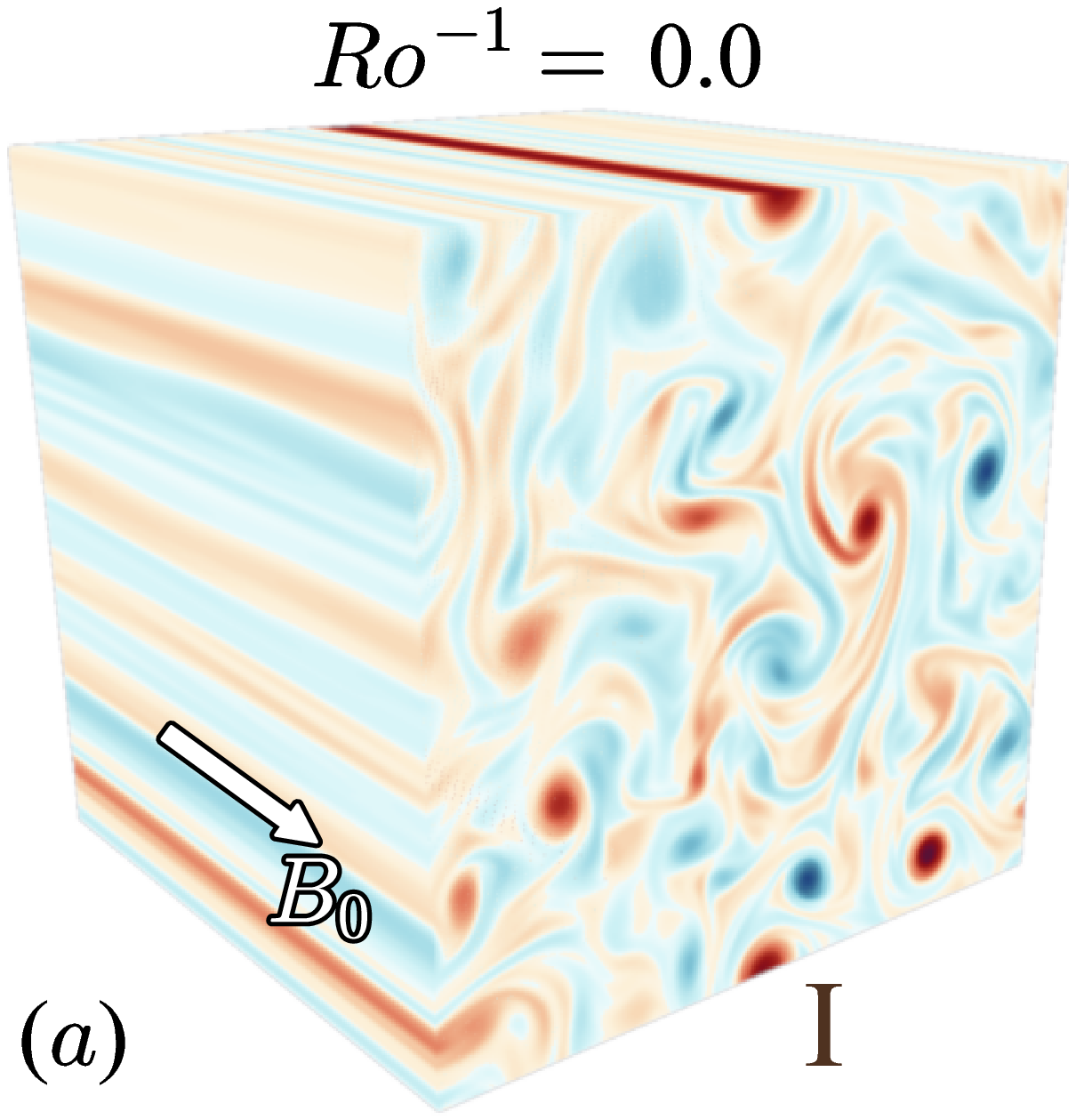}
		\includegraphics[width=0.48\textwidth]{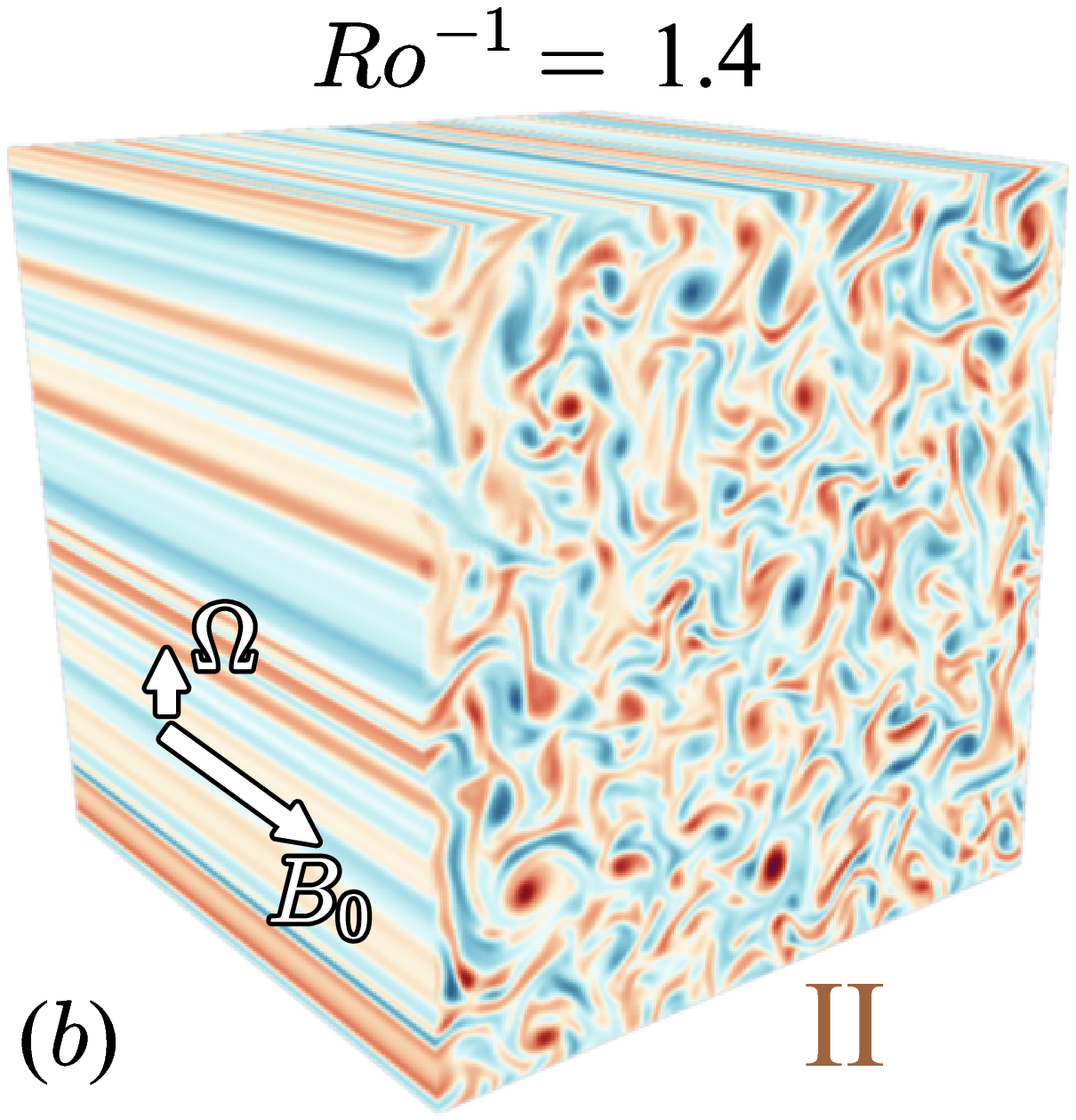}\\
		\includegraphics[width=0.48\textwidth]{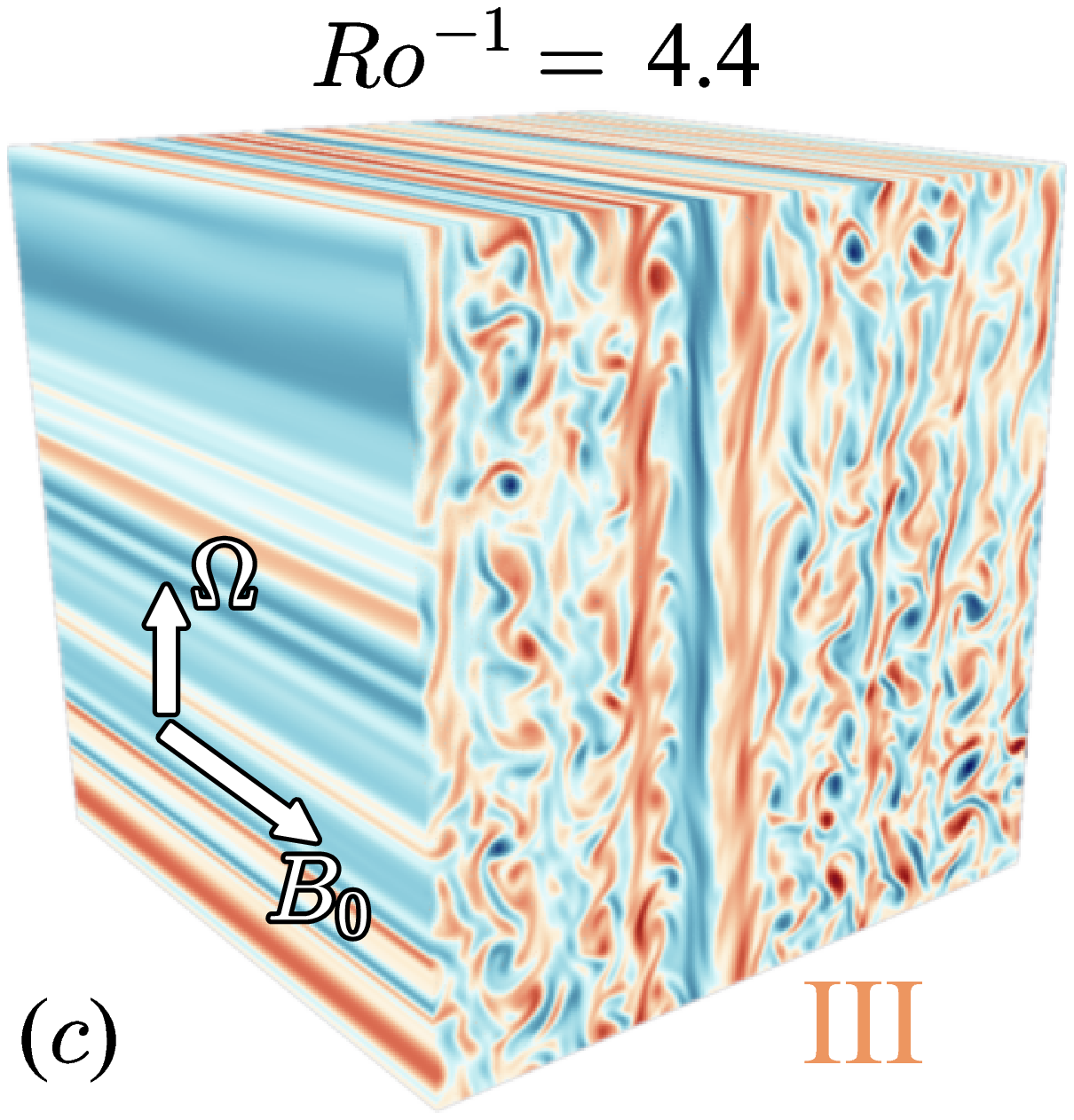}		\includegraphics[width=0.48\textwidth]{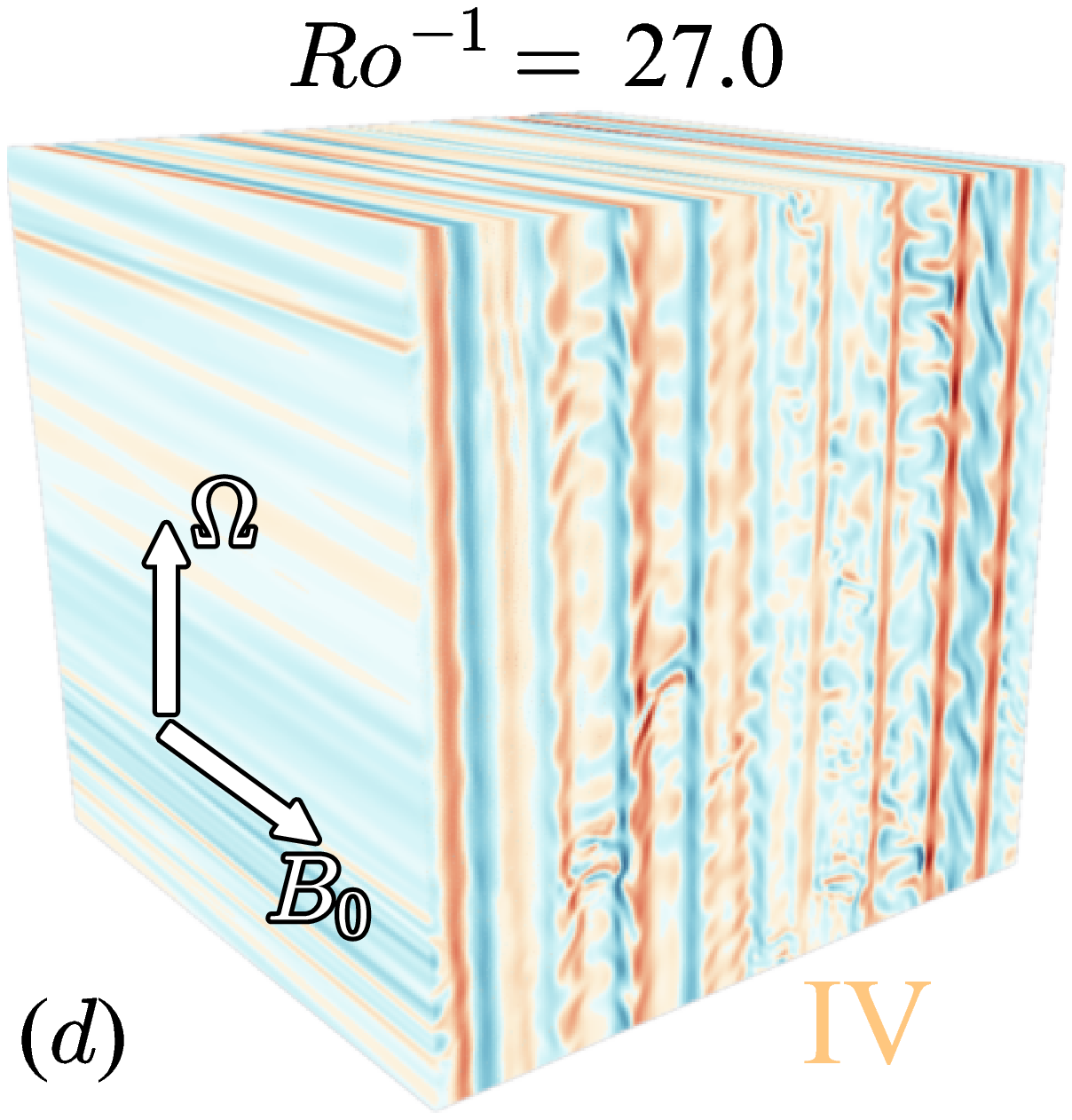}
	}
	\caption{
		Snapshots of the field-aligned vorticity $\omega = \hat{x}\cdot\left(\nabla \times \vv \right)$, representing, from top left to bottom right, Regimes I (\textit{a}), II (\textit{b}), III (\textit{c}), and IV (\textit{d}), as rotation rate is increased. The red colors represent positive vorticity whereas the blue represents negative vorticity. Regime I is characterized by a bidirectional cascade, Regime II a purely forward cascade, Regime III the formation of strong shear layers (seen here in the middle of the domain), and Regime IV magnetically active shear layers. Regimes I--III have a negligible induced magnetic energy, unlike Regime IV whose magnetic energy dominates the dynamics (Figure \ref{fig:diss3D}). }
	\label{fig:flow3D}
\end{figure}

All runs, unless otherwise stated, are in the large background field regime with $M^{-1} \approx 84$. 
We find this value to be large enough to produce the expected two-dimensionalization in the absence of rotation (Figure \ref{fig:flow3D}({\it a})). Larger background magnetic field values result in significant restrictions in the time-step which would limit our ability to perform the same parameter sweep.
The Reynolds and magnetic Reynolds numbers, defined, respectively, as $Re \equiv U / k_f^3 \nu$ and $Re_m \equiv U / k_f^3 \eta$ when considering hyperviscous and hyperdiffusive terms as we do, measure the relative strength of the advection terms compared to the hyperviscous and magnetic hyperdiffusion terms. For the simulations we performed, the Reynolds and magnetic Reynolds numbers were large (approximately 300) and equal to each other, i.e. the magnetic Prandtl number is set to one.  
We performed 14 runs at $M^{-1} \approx 84$ but at different values of $\OoRo$, ranging from $\OoRo = 0$ to $\OoRo = 27$. All averages and snapshots were taken in statistically steady states.
See Table \ref{tab:runs} for details of the simulations and a description of how we measured the nondimensional numbers. 

In this study, we are partly concerned with the behavior of the energy cascade as rotation is varied. We expect the presence of a bidirectional cascade, where a fraction of the energy input by the forcing goes to large scales and the rest goes to small scales. As such, we define a measure for the fraction of energy that goes to large scales in the form of kinetic energy, $\varepsilon_-$, and that which goes to small scales in the form of kinetic energy $\varepsilon$ and magnetic energy $\varepsilon_b$. Since the large-scale magnetic energy dissipation is practically zero for every simulation performed, we ignore it from our analysis, as it plays no role. These measures are based on the dissipation rates from each of the three dissipation terms, and are defined in the following way:
\begin{equation}
    \varepsilon_- \equiv \nu_- \langle \left| \nabla^{-1} \vv \right|^2 \rangle/I, \quad
    \varepsilon \equiv \nu \langle \left| \nabla^2 \vv \right|^2 \rangle/I, \quad
    \varepsilon_b \equiv \eta \langle \left| \nabla^2 \bb \right|^2 \rangle/I \label{eq:eps}.
\end{equation}
Energy balance at steady state tells us that $\varepsilon_- + \varepsilon + \varepsilon_b = 1$. 
In the limit of large Reynolds number and large forcing wavenumber, none of the energy injected is dissipated at the forcing scale and proper inertial ranges are formed. In this case, the dissipation rate at large scales represents the fraction of energy cascading to large scales, and similarly for the dissipation rate at small scales. Our runs do not reach these idealized limits.  The lack of scale separation between the forcing and large-scale dissipation will manifest itself in two related ways in this paper: (i) the large scale dissipation rate will remain nonzero despite zero average inverse cascade, because some energy that is being injected at $k_f$ will be dissipated by the large-scale dissipation mechanism ($\varepsilon_- \leq 0.1$ for the 3D runs), and (ii) when layers form in Regime III, both the 3D runs and 2D3C runs show an increase in large-scale dissipation rate, \textit{not} because of the presence of an inverse cascade, but because the layers form coherent structures near the forcing scale, their energy grows and hence a stronger large-scale dissipation rate is achieved. These jumps in $\varepsilon_-$ denote the presence of shear layers, as discussed in Section \ref{sec:2D}. To complement these estimates for energy cascades, we will look at the normalized spectral energy flux:
\begin{align}
    \Pi_{KE}(k) &\equiv \left\langle \vv^{<k} \cdot \left(\vv \cdot \nabla \vv \right)\right\rangle/I, \\
    \Pi_{ME}(k) &\equiv -\left\langle \vv^{<k} \cdot \left((\BB + \bb) \cdot \nabla \bb \right)\right\rangle/I + \left\langle \bb^{<k} \cdot \left(\vv \cdot \nabla \bb  - (\BB+\bb)\cdot \nabla \vv\right)\right\rangle/I,
\end{align}
where $\vv^{<k}$ stands for a filtering of the velocity $\vv$ in Fourier space so that only the wavenumbers with modulus smaller than $k$ are kept. 
The flux $\Pi(k)$ expresses the rate at which energy is flowing out of scales larger than $2\pi/k$ due to nonlinear interactions, normalized by the energy injection rate. Therefore, if energy is going from large to small scales, the energy flux will be positive, and \textit{vice versa}.
Finally, to quantify the amount and type of energy at each scale, we will also look at the energy spectra:
\begin{equation}
    E_{KE}(k) \equiv \frac{1}{2}\sum_{|\bv{k}|=k}|\widehat{\vv}|^2(\bv{k}), \quad E_{ME}(k) \equiv \frac{1}{2}\sum_{|\bv{k}|=k}|\widehat{\bb}|^2(\bv{k}),
\end{equation}
where $\widehat{\vv}$ denotes the Fourier transform of $\vv$.

Beginning from quasi-two-dimensional turbulence on the $y$-$z$ plane at zero rotation, we find four distinct regimes as we increased rotation (Figure \ref{fig:flow3D}). Although not so apparent in the 3D simulations, these regimes are separated by seemingly sharp transitions, whose boundaries are determined in section \ref{sec:2D}. 

\begin{figure}
	\centering{
		\includegraphics[width=0.9\textwidth]{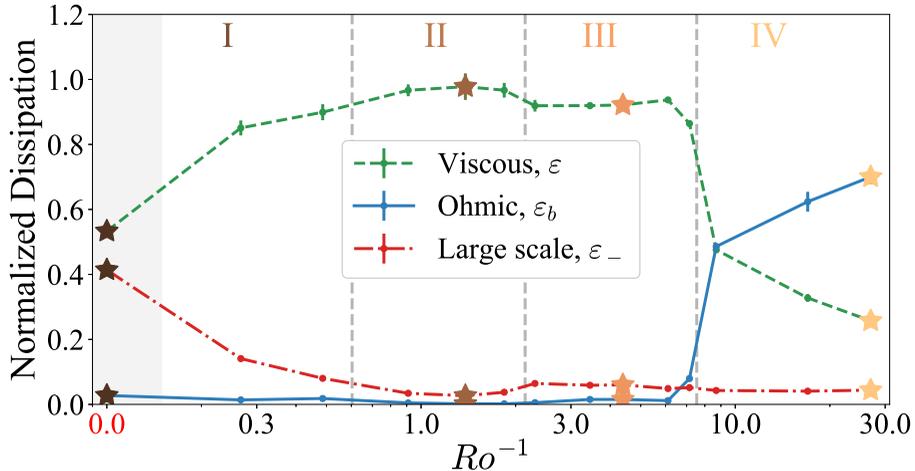}
	}
	\caption{Dissipation rates normalized by the energy injection rate as a function of rotation rate measured by the inverse Rossby number $\OoRo$. The blue solid line shows the Ohmic dissipation rate due to the magnetic diffusion term, $\varepsilon_b$, the green dashed line shows the viscous dissipation rate, $\varepsilon$, and the red dash-dotted line shows the large-scale dissipation rate due to the hypoviscosity, $\varepsilon_-$. Each regime is labeled at the top, and the vertical dashed lines represent boundaries between regimes, chosen based on the two-dimensional runs in section \ref{sec:2D}. Stars represent runs whose snapshots are shown in Figure \ref{fig:flow3D}.}
	\label{fig:diss3D}
\end{figure}

\begin{figure}
	\centering{
		\includegraphics[width=\textwidth]{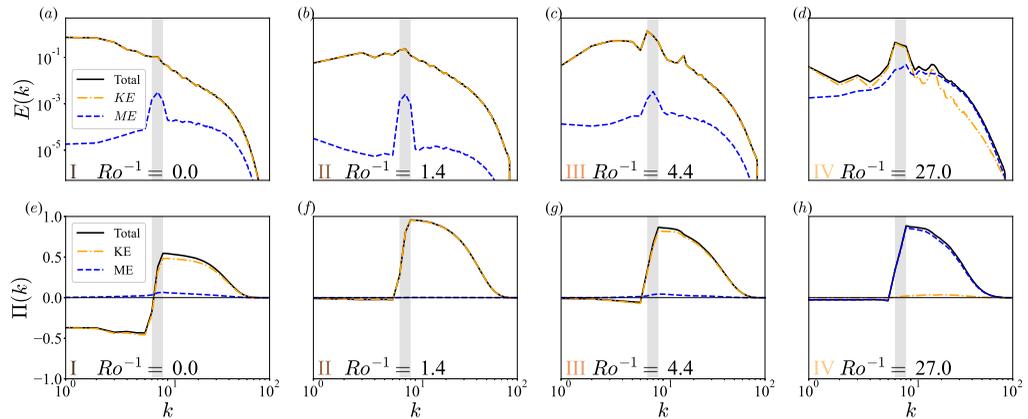}
	}
	\caption{The time-averaged energy spectra (top row) and spectral energy flux (bottom row) for each Regime found in our simulations. The blue dashed line shows the magnetic components, either $E_{ME}$ or $\Pi_{ME}$, the orange dash-dotted line shows the kinetic components, $E_{KE}$ or $\Pi_{KE}$, and the solid black line shows their sum. The grey box represents the forcing range. These are from the same simulations shown in Figure \ref{fig:flow3D} and which are starred in Figure \ref{fig:diss3D}. }
	\label{fig:spec}
\end{figure}

Regime I (Figure \ref{fig:flow3D}({\it a})), defined for runs with $\OoRo < 0.6$, is characterized by the presence of a bidirectional cascade. This can be seen in Figure \ref{fig:diss3D} as a non-zero large-scale dissipation rate as well as in Figure \ref{fig:spec}(\textit{e}), where the spectral energy transfers show that about half of the energy injected by the forcing goes to large scales (negative $\Pi(k)$) and the other half goes to small scales (positive $\Pi(k)$). 
The fraction of energy that goes to larger scales decreases with increasing rotation (Figure \ref{fig:diss3D}).
At zero rotation we don't have a purely inverse cascade ($\varepsilon_- \approx 1$) due to a combination of finite background magnetic field strength and, as we will see in section \ref{sec:2D}, the fact that we're forcing the out-of-plane velocity which acts as a passive scalar in the two-dimensionalized dynamics, thus contributing to a forward energy flux \citep{Campagne2014,Biferale2017}. Therefore, at zero rotation rate, the system is undergoing two independent cascades: an inverse energy cascade of horizontal kinetic energy and a forward cascade of the out-of-plane kinetic energy. If we were to force only the horizontal velocity components in the $k_x = 0$ wavenumber plane, we would expect to see $\varepsilon_- \approx 1$ at zero rotation. Figure $\ref{fig:spec}(\textit{a})$ shows the kinetic and magnetic energy spectra, which demonstrates that the magnetic energy is orders of magnitude smaller than the kinetic energy (particularly at large scales) and that the largest scales have the most energy, providing further confirmation of the presence of an inverse cascade. The spike of magnetic energy at the forcing scale is due to the excitation of Alfv\'{e}n waves from the isotropic forcing. The eddy length scales seen in Figure \ref{fig:flow3D}({\it a}) are set by a combination of the energy injection and the large-scale hypoviscosity coefficient.

Regime II (Figure \ref{fig:flow3D}(\textit{b})), defined for runs with $0.6 < \OoRo < 2.1$, is characterized by a purely forward cascade of energy (Figures \ref{fig:diss3D} and \ref{fig:spec}(\textit{f})). This may come as a surprise, given that the dynamics are two-dimensional. The reason for this seemingly-contradictory state is that, while two-dimensional, all three velocity components are active in the dynamics and, furthermore, are coupled together with rotation. This results in a set of reduced two-dimensional, three-component (2D3C) equations which no longer conserve enstrophy, making a forward cascade of energy possible. The rotating 2D3C system will be discussed and explored numerically in section \ref{sec:2D}.

Regime III (Figure \ref{fig:flow3D}(\textit{c})), defined for runs with $2.1 < \OoRo < 7.5$, is characterized by the formation of strong shear-layers along the $y$-direction, consisting of uniform velocity in the $x$-$z$ plane. The shear layers form when the rotational constraint on the dynamics at large scales becomes sufficiently large, requiring that $\partial_z \vv \approx 0$ at those scales. The combination of $\partial_z = \partial_x = 0$ and incompressibility implies that $v_y = 0$ (since we're in a periodic domain), and thus that the last remaining component of the nonlinear advection term $v_y \partial_y = 0$ and nonlinearities are suppressed at large scales.
Because of the suppressed nonlinearity at large scales, these shear layers form coherent structures that are fed directly by the forcing but that do not transfer that energy away, causing a build up of energy (not shown). The energy in the layers builds until a combination of the large-scale dissipation (Figure \ref{fig:diss3D}) and the nonlinear term (Figure \ref{fig:spec}(\textit{g})) are able to remove energy from those scales. Regimes I--III have negligible induced magnetic energy, as is observed in simulations of MHD with a strong background field \citep{Alexakis2011,Sujovolsky2016}, and so the induced magnetic field plays an insignificant role in the dynamics. The magnetic fluctuations are also much smaller than $B_0$ -- less than 0.5\% of $B_0$ in Regimes I-III.

This changes, however, in Regime IV (Figure \ref{fig:flow3D}(\textit{d})), defined for runs with $\OoRo > 7.5$, where we have found the activation and growth of the induced magnetic field, which dominates both the energy as well as the nonlinear energy transfers (Figure \ref{fig:diss3D}). The nonlinear advection term in the momentum equation is suppressed for practically all scales (Figure 3(\textit{h})), leading to laminar-like shear-layer structures (Figure \ref{fig:flow3D}(\textit{d})) and a turbulent magnetic field which is responsible for the nonlinear transfers of energy across scales, via the Lorentz force and the magnetic induction equation. The shear layer spacing in Figure \ref{fig:flow3D}({\it d}) is set by the forcing scale. In this regime, significant induced magnetic field fluctuations occur both parallel and perpendicular to the background magnetic field, with a magnitude of about 3\% of $B_0$.

We expect the boundaries between Regimes I--III to be independent of $M^{-1}$, as they are part of the asymptotic 2D3C HD, whose sole parameter is the rotation rate. We confirm this in the next section, which deals specifically with this asymptotic set of equations, by showing that the regime transitions happen for the same values of $\OoRo$.
The transition from Regime III to IV is of a different nature and represents a breakdown of the hydrodynamic behavior found for lower rotation rates. This transition is found to be $M^{-1}$-dependent, and will be discussed briefly in section \ref{sec:conclusions}.

\begin{table}
	\begin{center}
		\def~{\hphantom{0}}
		\begin{tabular}{ccccccccccc}
		 System & $B_0$ & $M^{-1}$ & $\Omega$   & $k_f$ & Forcing type & $\nu$ & $\nu_-$ & Resolution & Count \\[3pt]
		(\ref{eq:RMHDB1})--(\ref{eq:RMHDB3}) & 13.3 & 84 &  [0 $-$ 16.6] & 8-10 &  Constant Amplitude  & 6.6e-7  & 0.06 &$256^3$ & 14 \\
		(\ref{eq:RMHDB1})--(\ref{eq:RMHDB3}) & 6.6 & 44 &  [0 $-$ 7.3] & 8-10 &  Constant Amplitude   & 6.6e-7  & 0.06 &$256^3$ & 11 \\
		(\ref{eq:RMHDB1})--(\ref{eq:RMHDB3}) & 3.3 & 23 &  [0 $-$ 7.3] & 8-10 &   Constant Amplitude  & 6.6e-7  & 0.06 &$256^3$ & 5 \\ 
		(\ref{eq:2D3C1})--(\ref{eq:2D3C2})  & $\infty$ & $\infty$ &  [0 $-$ 20] & 12  &   Random & 4e-7  & 1.0   & $512^2$ & 23 \\ 
		\end{tabular}
		\caption{A summary of the runs performed for this work. All runs have hyper- and hypo-viscosity of the same order (section \ref{sec:3D}). For runs with a magnetic field, $\mu = \nu$ and $\mu_- = \nu_-$. The simulations used Alfv\'{e}nic units so that $B_0 / \sqrt{\rho_0 \mu_0} \rightarrow B_0$ and the other $\rho_0$ was absorbed into the pressure. The values are non-dimensionalized by $L$ and the forcing amplitude $f_0$ (or $I_0$ for random forcing), so that $k_f$ is the forcing mode number and the forcing amplitude (for constant amplitude forcing) or energy injection rate (for the random forcing) are both 1 in these units. The typical velocity, $U$, was calculated after-the-fact for each run using $U^3 \equiv Ik_f^{-1}$, where $I \equiv \langle \bv{f} \cdot \vv \rangle$ is the time- and space-averaged energy injection rate. The count is the number of runs in that set.}
		\label{tab:runs}
	\end{center}
\end{table}


\section{Comparison to rotating two-dimensional, three-component model}\label{sec:2D}
Regimes I--III can be better understood by considering the asymptotic limit of equations (\ref{eq:RMHDB1}) -- (\ref{eq:RMHDB3}) when taking $M^{-1} \rightarrow \infty$ and keeping $\OoRo \sim \mathcal{O}(1)$ and the domain size fixed. The choice of keeping the domain size fixed is based on the fact that we are motivated mostly by astrophysical settings in confined geometries, in the presence of a strong background magnetic field. We acknowledge, however, that in many other astrophysical settings, such as the extended atmosphere of stars or the interstellar medium, a confined geometry may not be the best representative system to study. In such systems, a more appropriate limit might include taking the domain size to infinity at the same time as the $M^{-1}$, so as to prevent the exact two-dimensionalization of the flow \citep{Thess2007,GalletDoering2015}. The limiting equations in this case would resemble more the Reduced MHD system, derived for tokamaks but used also to study some astrophysical systems, in which the flow is highly anisotropic, yet still three-dimensional \citep{Strauss1976,Oughton2017}. 

Our limiting procedure, with fixed domain size, is similar to that done in \cite{Montgomery1981}, with the exception that we include the Coriolis term, and so we only briefly discuss it here. Assuming a background magnetic field in the $x$ direction and a rotation axis along the $z$ direction, the process results in a set of three dynamical equations and one nonlinear constraint for the three variables: $\psi(y,z,t)$ the streamfunction for the in-plane velocities, $v_x(y,z,t)$ the out-of-plane velocity, and $A(y,z,t)$ the potential for the in-plane magnetic field.  This novel constraint, which results from the presence of the Coriolis term, states that either $\delta A / \delta v_x =0$ or $v_y = \partial_z \psi = 0$, where the former is the functional derivative of $A$ with respect to $v_x$. Our three-dimensional simulations from section \ref{sec:3D} seem to be consistent with these constraints, where, in Regimes I--III for $\OoRo < 7.5$, we have $A\approx 0$ but $v_y \neq 0$ and, in Regime IV for $\OoRo > 7.5$, we have $A\neq 0$ but $v_y = \partial_z \psi \approx 0$. For the purposes of studying the reduced dynamics of Regimes I--III, we assume $A = 0$, knowing that it would not capture the transition to Regime IV. The resulting equations form the two-dimensional, three-component (2D3C) system with in-plane rotation:
\begin{align}
	\frac{\D v_x}{\D t} + [v_x,\psi] &= 2 \Omega \frac{\D \psi}{\D z} + \nu \nabla_\perp^2 v_x + f_x, \label{eq:2D3C1} \\
	\frac{\D \omega}{\D t} + [\omega,\psi] &= 2 \Omega \frac{\D v_x}{\D z} + \nu \nabla_\perp^2 \omega + f_\omega, \label{eq:2D3C2}
\end{align}
where $[F,G] \equiv \partial_y F \partial_z G - \partial_y G \partial_z F = 0$, $\nabla_\perp = (0,\partial_y,\partial_z)$, $\omega = \hat{x}\cdot (\nabla \times \vv ) = -\nabla^2_\perp \psi$ is the out-of-plane vorticity (of the in-plane velocities), $\perp$ implies the directions perpendicular to the background magnetic field, and $f_\omega = \hat{x}\cdot (\nabla_\perp \times \bv{f}_\perp)$. One could equivalently arrive at (\ref{eq:2D3C1}) and (\ref{eq:2D3C2}) by taking rotating 3D HD and requiring that the velocity field doesn't depend on $x$. If considering an arbitrary angle between the background field and rotation, only the perpendicular projection of the rotation vector on the background field enters the model. For example, supposing without loss of generality that $\BB = B_0 \bv{\hat{x}}$ and $\OO = \Omega (\sin(\theta) \bv{\hat{z}} + \cos(\theta) \bv{\hat{x}})$, then the Coriolis terms on the right-hand-side of (\ref{eq:2D3C1}) and (\ref{eq:2D3C2}) will be multiplied by $\sin(\theta)$. This asymptotic model is in the same spirit as some of the magnetized quasigeostrophic models used in astrophysical applications \citep{Aurnou2015,Maffei2019}, but it is important to note that here we have taken $M^{-1} \rightarrow \infty$ while keeping $\OoRo \sim \mathcal{O}(1)$, whereas the magnetized quasigeostrophic models take the rapidly rotating limit first. As is discussed in section \ref{sec:conclusions}, these limits are not expected to be the same.

The Coriolis force now couples the two equations together, making what would otherwise be a passive tracer into an active one. In fact, for non-zero rotation, it can be shown that the 2D3C rotating system conserves kinetic energy and helicity:
\begin{align}
    KE &= \frac{1}{2}\int v_x^2 + |\nabla \psi|^2 \ d^2 x, \\
    H &= \int v_x \omega \ d^2 x.
\end{align}
These are the same conserved quantities as in 3D HD, but we emphasize that the dynamics are two-dimensional and are occurring on the $y$-$z$ plane.
This is in contrast to the case of zero rotation, where the system conserves (separately) the in-plane kinetic energy $\int |\nabla \psi|^2 \ d^2x$ and the out-of-plane kinetic energy $\int v_x^2 \ d^2x$, as well as the enstrophy, $\int \omega^2 \ d^2 x$. The conservation of enstrophy can be shown to prevent the existence of a forward cascade of in-plane kinetic energy \citep{Fjortoft1953,Kraichnan1967,Alexakis_Review}. Without the restriction of enstrophy conservation, though, the kinetic energy may go downscale in a forward cascade, even if one doesn't force the out-of-plane component.

We solve equations (\ref{eq:2D3C1}) and (\ref{eq:2D3C2}), with modified hyper- and hypo-viscosities as in the 3D simulations, numerically in a doubly-periodic domain of side length $2 \pi L$ using the 2D predecessor of GHOST. The code can be found in a public Github repository \citep{Benavides_Code_2D}. Unlike the 3D runs, whose forcing function had a constant amplitude in time, the 2D3C runs have random (white-in-time) forcing. At each time step, a wavenumber $\bv{k}_r$ of magnitude $k_f$ is chosen at random, and $\hat{f}_\omega (\bv{k})$ (Fourier transform of $f_\omega$) is set to zero everywhere except for at $\bv{k}_r$, where it had a magnitude $k_f \sqrt{2 I_0 / \Delta t}$ \citep{Chan2012}. This has the effect of setting the energy injection rate for the in-plane flow to be $I = \langle \psi f_\omega \rangle = I_0$ on average, with $I_0$ being an input parameter of the simulation. The same forcing is applied for $f_x$, but with an amplitude of $\sqrt{2 I_0 / \Delta t}$ instead, giving the same results. Therefore, half of the energy is injected into the in-plane flow and the other half in the out-of-plane velocity. We nondimensionalize all dynamical variables as before, using $L$ and now the energy injection rate parameter $I_0$. For all of the runs reported $k_f = 12$. See Table \ref{tab:runs} for details on the runs.
\begin{figure}
	\centering{
		\includegraphics[width=0.32\textwidth]{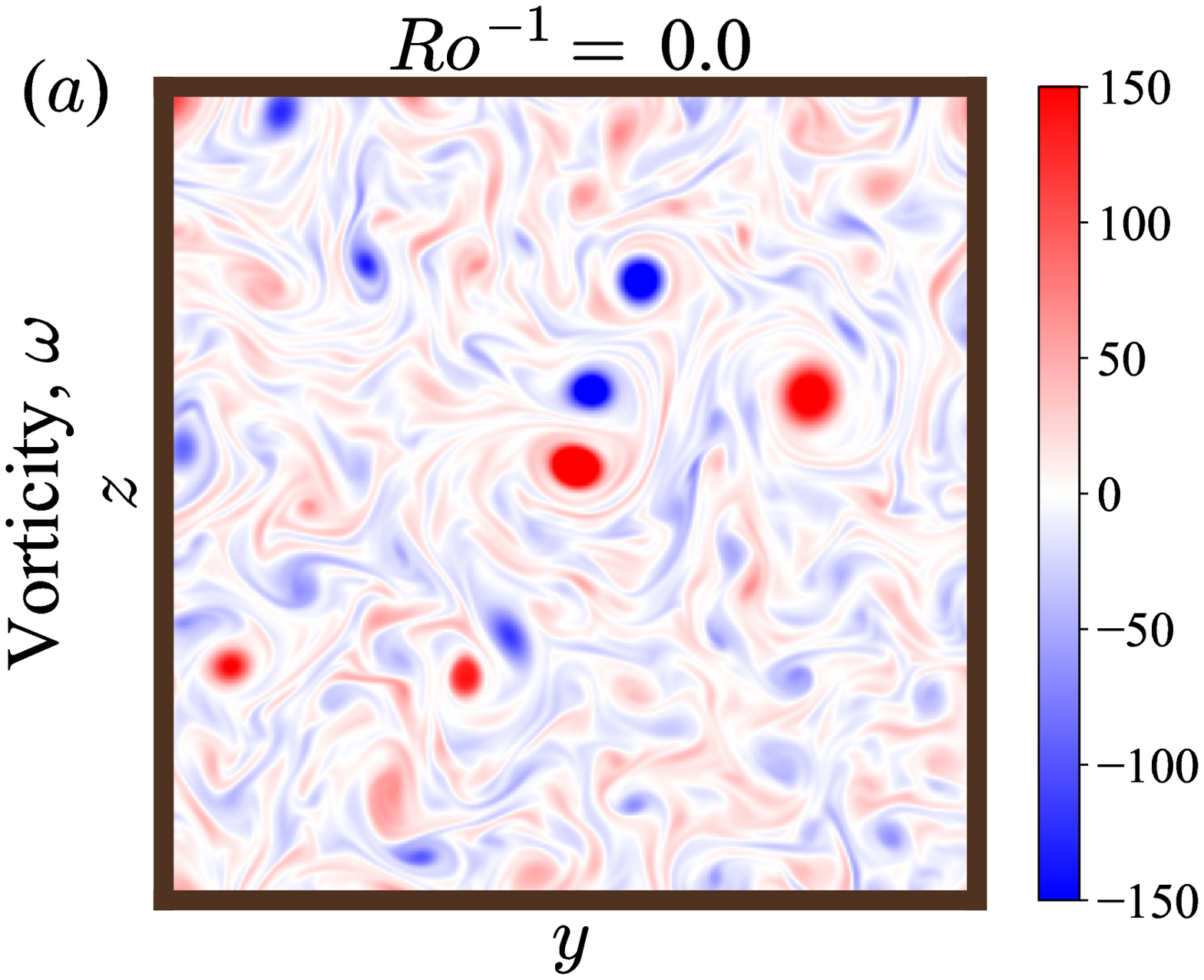}
		\includegraphics[width=0.32\textwidth]{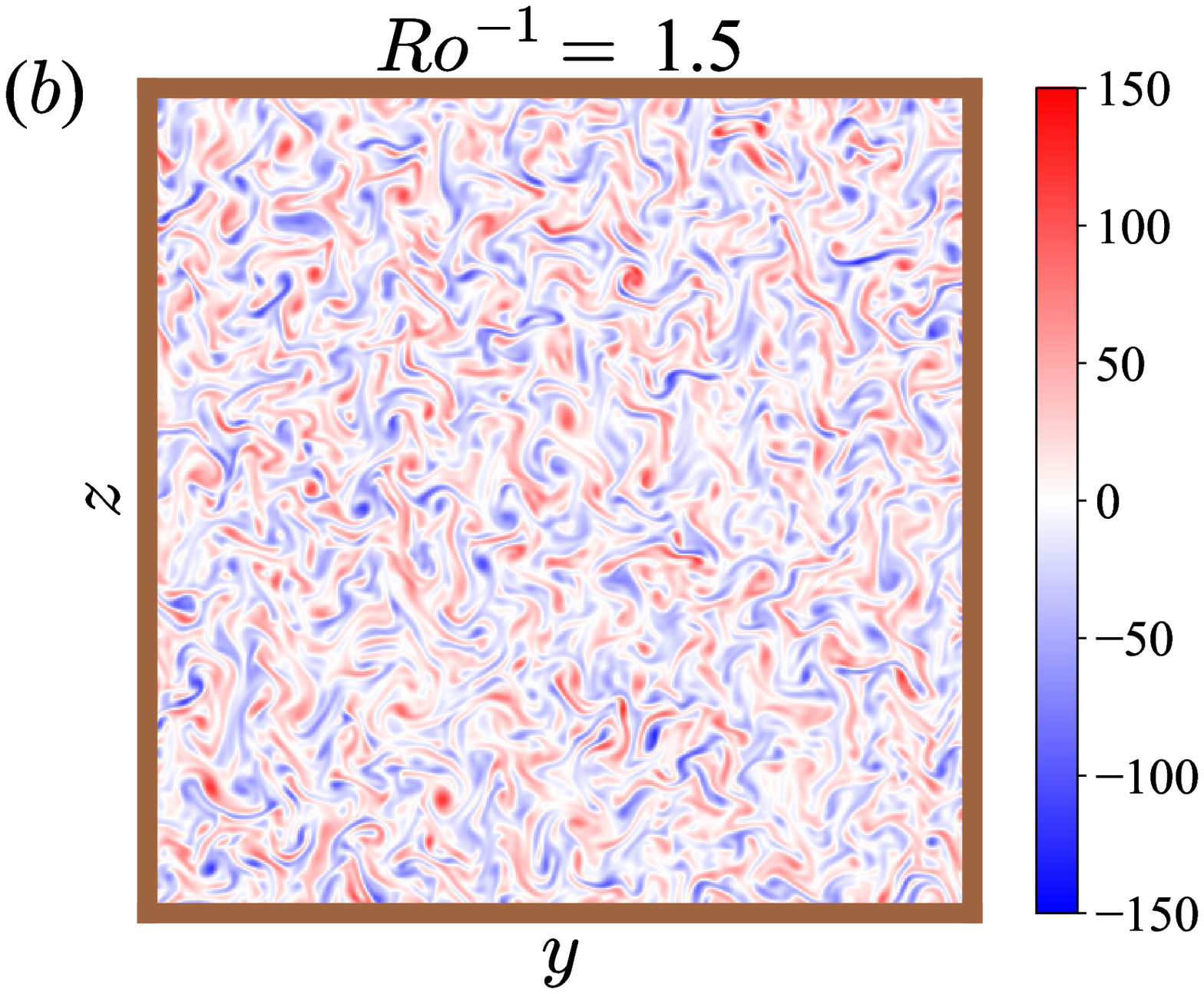}		\includegraphics[width=0.32\textwidth]{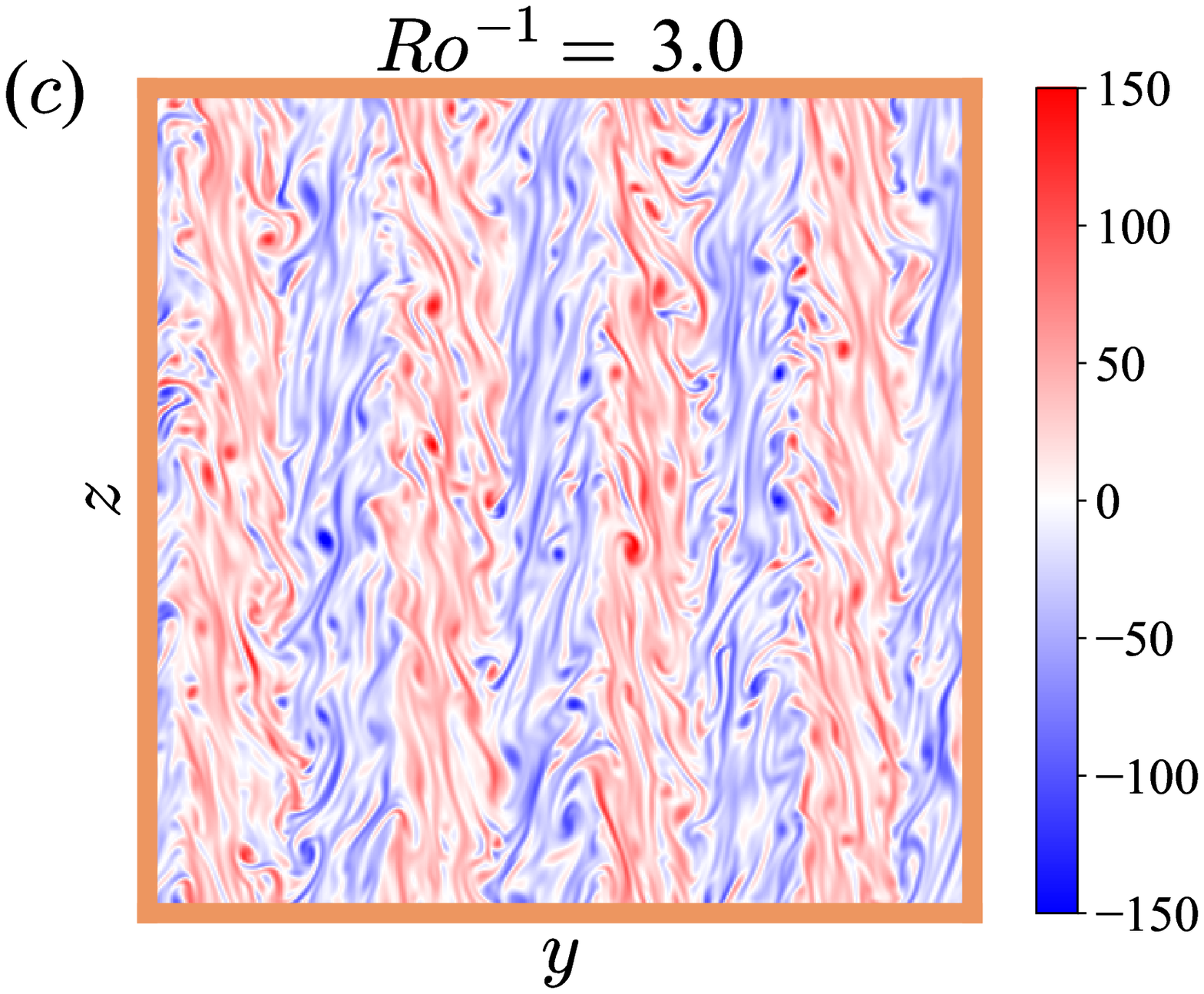}
	}
	\caption{
		Snapshots of the out-of-plane vorticity, $\omega = \hat{x}\cdot (\nabla \times \vv ) = -\nabla^2_\perp \psi$, for the 2D3C rotating simulations, representing, from left to right, Regimes I (\textit{a}), II (\textit{b}), and III (\textit{c}), as rotation is increased. We see striking similarities to Figure \ref{fig:flow3D}, confirming that the asymptotic 2D3C model captures the different regimes found in the 3D simulations of section \ref{sec:3D}.}
	\label{fig:flow2D}
\end{figure}

The goal of these simulations is to reproduce the parameter sweep performed in section \ref{sec:3D}, but with the added advantage of working with a two-dimensional code, thus allowing a larger quantity of runs, higher resolutions (larger Reynolds numbers, around 600), and longer time integration. We have performed 23 runs, with $\OoRo$ ranging from 0 to about 5, at four times the horizontal resolution. Our results confirm the presence of Regimes I--III, going from a bidirectional cascade to a forward cascade to a shear-layer configuration (Figure \ref{fig:flow2D}). 

\begin{figure}
	\centering{
		\includegraphics[width=0.9\textwidth]{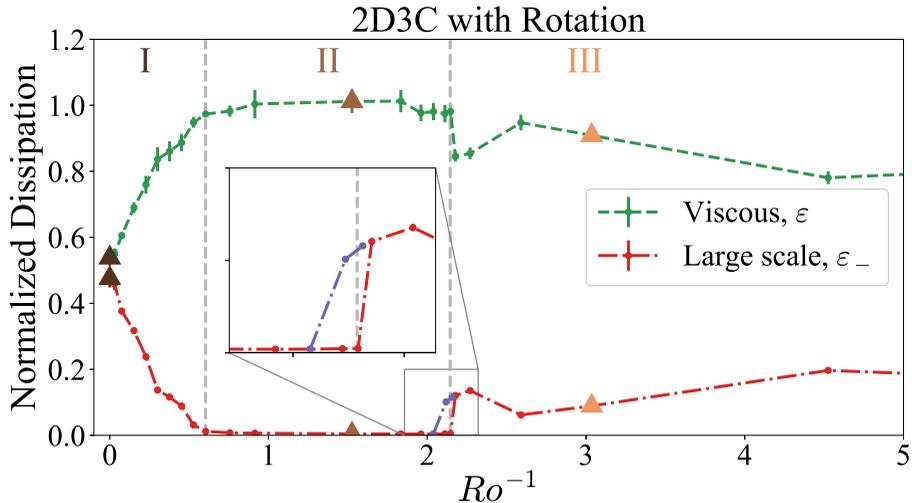}
	}
	\caption{Dissipation rates normalized by the energy injection rate as a function of rotation rate measured by the inverse Rossby number $\OoRo$. The green dashed line shows the viscous dissipation rate, $\varepsilon$, the red dash-dotted line shows the large-scale dissipation rate due to the hypoviscosity, $\varepsilon_-$, and the purple dash-dotted line shows the same but for hysteresis runs initialized with layers. Each regime is labeled at the top, and the vertical dashed lines represent boundaries between regimes. These denoted boundaries are placed at the same value of $\OoRo$ as those seen in Figure \ref{fig:diss3D}. Triangles represent runs whose snapshots are shown in Figure \ref{fig:flow2D}.}
	\label{fig:diss2D}
\end{figure}

At zero rotation we see a bidirectional cascade with half of the injected energy going to large scales and half going to small scales (Figure \ref{fig:diss2D}), similar to what was found in the 3D runs (Figure \ref{fig:diss3D}). For the 2D3C rotating system this is the case because of the choice of forcing, which injects half of the energy to the in-plane flow and the other half to the out-of-plane velocity. Since the two flows are completely decoupled at zero rotation, they each follow the standard behavior observed in 2D and passive tracer turbulence, that is, an inverse and forward cascade of energy, respectively. 
As we increase rotation, the Coriolis force couples the two fields, enstrophy is no longer conserved, and the in-plane velocities no longer cascade all the injected energy to large scales, resulting in a bidirectional cascade with decreasing inverse energy flux. There is an approximately linear approach to zero inverse energy flux, and at $\OoRo \approx 0.6$ there is a transition to a purely forward cascade. 
With a larger number of simulations, Regimes I and II are much more clearly separated, and their transition appears to be sharp (Figure \ref{fig:diss2D}).
This transition is qualitatively similar to other bidirectional to forward cascade transitions seen in other studies and could hint at a universal mechanism responsible for this transition \citep{Seshasayanan2014,Seshasayanan2016,Benavides2017,VanKan2020}.

Upon further increase of the rotation, the forward cascade regime (Regime II) transitions to a shear layer configuration (Figure \ref{fig:flow2D}(\textit{c})), entering Regime III.
This corresponds to the case when the Coriolis force dominates at large scales, making the dominant balance in equations (\ref{eq:2D3C1}) and (\ref{eq:2D3C2}) $\partial_z \psi \approx \partial_z v_x \approx 0$, hence the layers. There are a few differences in the morphology of the shear layers seen for these runs, compared to Regime III in the 3D simulations (Figure \ref{fig:flow3D}(\textit{c})). Here they take up the whole domain and also appear to equilibrate at scales larger than the forcing, through a series of mergers (not shown). Neither of these characteristics are seen in the shear layers of the 3D simulations. We believe this is due to a few factors, including the longer integration times and the change in forcing. 
A surprising feature of this transition, revealed by the better-resolved parameter sweep, is that it is discontinuous (Figure \ref{fig:diss2D}) 
\footnote{An increase in large-scale dissipation marks this transition \textit{not} because an inverse cascade forms (a weakness of this measure), but because of a lack of separation of scales. The layers form at or near the forcing scale and remain there as coherent structures, fed directly by the forcing, resulting in a build up of energy at those scales. This, in turn, results in a larger dissipation rate from the large-scale dissipation. If we were to perform runs at a larger $k_f$, this effect would disappear. The discontinuous transition is also observed in the kinetic energy, which is not shown.}.
Discontinuities are a characteristic of subcritical bifurcations, which should also display hysteresis. By initializing in the layered regime, we confirmed the presence of hysteresis as we reduced the rotation rate (Figure \ref{fig:diss2D} inset).

Despite differences in the forcing, Reynolds numbers, and values of $M^{-1}$, the regime transitions seem to occur for the same values of $\OoRo$, suggesting that the rotating 2D3C system successfully describes the dynamics observed in the 3D simulations from section \ref{sec:3D} and that Regimes I--III are robust properties of the system. The two-dimensional asymptotic system has allowed us to perform a more detailed parameter sweep of this parameter space, and has revealed sharp transitions and nontrivial behavior near those transitions which we did not anticipate from the 3D simulations. 


\section{Discussion \& Conclusions}\label{sec:conclusions}
We have investigated the turbulent dynamics of rotating magnetohydrodynamics in the presence of a strong uniform background magnetic field perpendicular to the rotation axis. Our investigations have revealed surprising behavior, confirmed both by three-dimensional and a two-dimensional three-component asymptotic model, as rotation rate is increased. We observed the weakening of the inverse cascade, a transition to a purely forward cascade for relatively weak rotation, and eventually a shear-layer regime at larger rotation rates. 
These results were obtained in a specific situation: orthogonal rotation and guide field at unit magnetic Prandtl number. However, the derivation of the asymptotic 2D3C model allows us to generalise Regimes I-III to the more realistic situation of an arbitrary angle between rotation and guide field at low magnetic Reynolds number. First, for an arbitrary angle $\theta$ between rotation vector and guide field, the reduced model is given by equations (\ref{eq:2D3C1}) and (\ref{eq:2D3C2}) where $2 \Omega$ is replaced by $2 \Omega \sin (\theta)$, the consequence being that the results in Figure \ref{fig:diss2D} carry over with $Ro^{-1}$ replaced by $Ro^{-1} \sin (\theta)$ in the $x$-axis. Second, the 2D3C model illustrates the asymptotic limit in which the guide field is so strong that it prevents any $x$-dependence. The same phenomenon arises for the low magnetic Reynolds numbers that characterize transitions regions in planetary interiors, see \cite{GalletDoering2015} for a rigorous proof in an idealized setting. The consequence is that we expect the very same reduced 2D3C model to hold at low magnetic Reynolds number, starting either from the full MHD equations or from their low-magnetic-Reynolds-number quasi-static approximation. We thus conclude that Regimes I-III carry over to the planetary relevant situation of an arbitrary angle between rotation and guide field, together with a low magnetic Reynolds number (by contrast, the magnetically active Regime IV will be affected by changes in magnetic Prandtl number).

We should also stress the fact that our study focuses on finite-size domains: motivated by transitional layers in planetary interiors, we have restricted attention to a numerical domain that is finite both along the direction of the rotation vector and the local direction of the large-scale magnetic field. By contrast, an idealized turbulent cloud allowed to develop arbitrarily large structures would never achieve exact two-dimensionalization \citep{Davidson2013,VanKan2020,VanKan2021}, and it's possible that in those cases the Reduced MHD description might be more relevant \citep{Strauss1976,Oughton2017}.

The strong sensitivity of the inverse cascade to in-plane rotation could have significant implications for the morphology of astrophysical flows, which often have both rotation and a background magnetic field. Even for relatively weak rotation ($\OoRo \sim 1$) the inverse cascade is entirely suppressed. Seeing as an inverse cascade is considered to be necessary for the formation of jets on gas giant planets, this phenomenon could be a tentative alternative explanation for the weakening of the jets in the depths of their atmospheres, as seen by the \textit{Juno} mission on Jupiter \citep{Kaspi2018}. In the outer electrically-neutral regions the jets can form because of the rapid rotation. These rotation-aligned jets may penetrate deep into the interior until they reach the low $Re_m$ ionized regions of the atmosphere whose turbulent dynamics suppress the jets  via Ohmic dissipation \citep{Liu2008}. Our work reveals another potential alternative, where a misalignment of the rotation and background field cause the localized turbulent dynamics to cascade energy forward instead of inversely, thereby taking away the dynamical origin of the jets. Apart from the astrophysical implications, the rotating 2D3C model might be of interest to those studying phase transitions in turbulence \citep{Alexakis_Review} -- particularly those interested in the transition from a forward to a bidirectional cascade, since, as far as we are aware, this model is the only two-dimensional hydrodynamical one with this behavior.
\begin{figure}
	\centering{
		\includegraphics[width=0.9\textwidth]{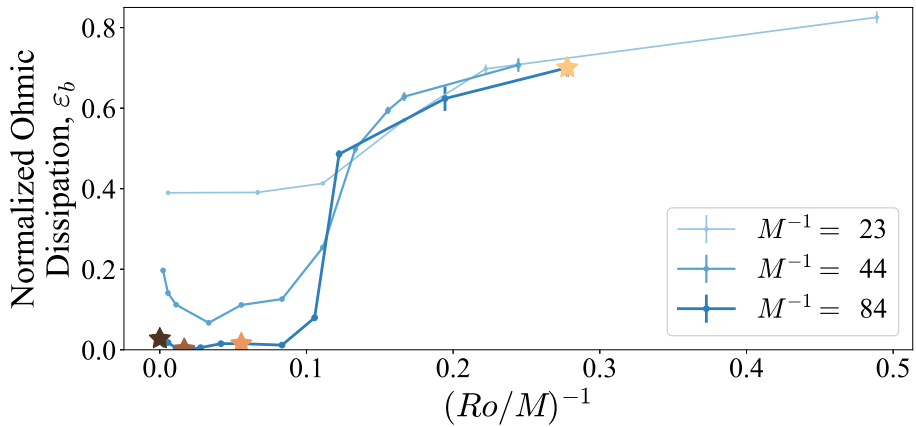}
	}
	\caption{Normalized Ohmic dissipation, $\varepsilon_b$, versus $(Ro/M)^{-1}$. The Ohmic dissipation represents a measure of how active the induced magnetic field is in the dynamics. We see that, for three values of $M^{-1}$, the induced magnetic field begins to dominate the dynamics once $(Ro/M)^{-1} > 0.1$, in other words when $\OoRo > 0.1 M^{-1}$. $(Ro/M)^{-1}$ is also referred to as the Lehnert number \citep{Lehnert1955}. Stars represent runs whose snapshots are shown in Figure \ref{fig:flow3D}.}
	\label{fig:odiss}
\end{figure}

At the largest rotation rotates, our 3D simulations showed a sudden activation of the induced magnetic field, signaling the breakdown of the purely hydrodynamic asymptotic model. The velocity field remained 2D3C, but the dynamics differed significantly from the hydrodynamic shear-layer regime and were dominated by an induced three-dimensional magnetic field. Although the 2D3C model breaks down, given the three-dimensionality of the magnetic field, it's possible this transition could be studied with the Reduced MHD system \citep{Strauss1976,Oughton2017}. A series of simulations at lower $M^{-1}$ values (Table \ref{tab:runs}) reveal that the transition happens when $M \sim Ro$, which represents roughly the point at which the inertial wave frequency begins to dominate over the Alfv\'{e}n wave frequency (Figure \ref{fig:odiss}, Appendix \ref{app:waves}). Interestingly, this transition sharpens towards a critical value as the background magnetic field strength increases. Therefore, when considering the limit of strong rotation and strong background magnetic field, the order in which those limits are taken matters. If $\OoRo < 0.1 M^{-1}$, then one would expect a hydrodynamical regime, whereas if $\OoRo > 0.1 M^{-1}$ a magnetically-dominated regime is expected.

\section*{Acknowledgments}
The authors thank the referees for helpful feedback which has improved the quality and clarity of the manuscript. 

\section*{Funding}
This research was carried out in part during the 2019 Summer School at the Center for Computational Astrophysics, Flatiron Institute. The Flatiron Institute is supported by the Simons Foundation.
SJB acknowledges funding from a grant from the National Science Foundation (OCE-1459702) and from the National Aeronautics and Space Administration (Award Number: 80NSSC20K1367) issued through the Future Investigators in NASA Earth and Space Science and Technology (NNH19ZDA001N-FINESST) within the NASA Research Announcement (NRA): Research Opportunities in Space and Earth Sciences (ROSES-2019).

\section*{Declaration of Interests}
The authors report no conflict of interest.

\section*{Data availability statement}
The three-dimensional $B\Omega$-MHD runs were done using \cite{Benavides_Code}. The two-dimensional 2D3C runs were done using \cite{Benavides_Code_2D}. The time-averaged data, which includes run parameters, and a script for creating the figures can be found at \url{https://doi.org/10.6084/m9.figshare.16888135.v1}.
The time-series used to create this data can be provided upon request to the corresponding author.

\appendix
\section{Wave Dispersion Relation}\label{app:waves} 
\begin{figure}
	\centering{
		\includegraphics[width=\textwidth]{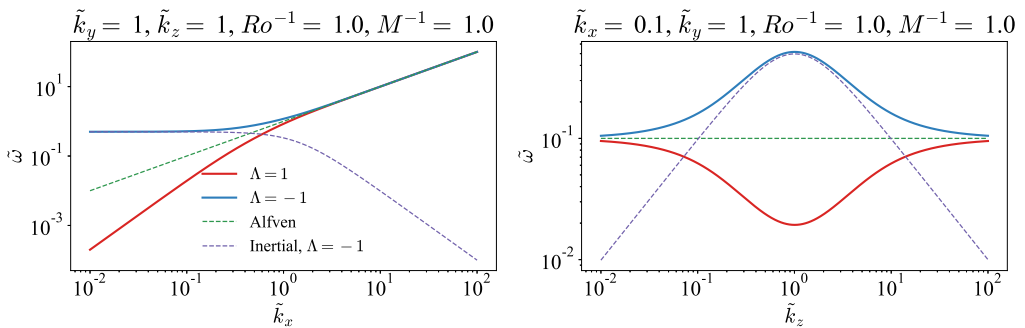}
	}
	\caption{The dispersion relation for waves in the $B\Omega$-MHD system, in the case where $\BB = B_0 \bv{\hat{x}}$ and $\OO = \Omega \bv{\hat{z}}$ (equation (\ref{eq:disprel2})). The green dashed line shows the Alfv\'{e}n wave dispersion relation, and the purple dashed line shows that of an inertial wave. The low frequency branch with $\Lambda = 1$ is similar to the magnetostrophic mode in the case when $\BB$ and $\OO$ are aligned \citep{Galtier2014}.}
	\label{fig:waves}
\end{figure}

In the inviscid, perfectly-conducting, and force-free case, the linearized $B\Omega$-MHD system admits wave solutions. Taking $\vv$ of the form $\vv = \hat{\vv} \exp{i(\bv{k}\cdot \bv{x} - \omega t)}$ and plugging this into the linearized versions of equations (\ref{eq:RMHDB1})-(\ref{eq:RMHDB3}), after some algebra we end up with:
\begin{equation}
    \omega^2 \left(\bv{k} \times \hat{\vv} \right) = 2 i \omega \left(\bv{k}\cdot \OO\right) \hat{\vv} + \left(\bv{k} \cdot \BB\right)^2 \bv{k} \times \hat{\vv}. \label{eq:disprel1}
\end{equation}
Next we introduce the helical orthonormal basis for $\hat{\vv}$, $\hat{\vv}(\bv{k}) = v_k^+ \hat{\bv{h}}_k^+ + v_k^- \hat{\bv{h}}_k^- $, where $\bv{k} \times \hat{\bv{h}}_k^\Lambda = -i \Lambda |\bv{k}|\hat{\bv{h}}_k^\Lambda$ and $\Lambda = \pm 1$ indicates the sign of the helicity of $\hat{\bv{h}}_k^\Lambda$ \citep{Herring1974,Alexakis2017}. Introducing these basis and dotting equation (\ref{eq:disprel1}) with $\hat{\bv{h}}_k^\Lambda$ we arrive at the dispersion relation. We normalize the frequency with the eddy turnover frequency, $k_f U$, and the wavevector $\bv{k}$ with $k_f$, resulting in our final expression for the dispersion relation:
\begin{equation}
    \tilde{\omega}\left(\bv{\tilde{k}}; \Lambda\right) = -\frac{1}{2}\frac{\bv{\tilde{k}}\cdot \opara}{\Lambda \tilde{k} Ro} \pm \frac{1}{2}\sqrt{\left(\frac{\bv{\tilde{k}}\cdot \opara}{\tilde{k} Ro}\right)^2 + 4 \left(\frac{\tilde{\bv{k}}\cdot \bpara}{M}\right)^2},
\end{equation}
where  $\tilde{k} \equiv (\tilde{k}_x^2 + \tilde{k}_y^2 + \tilde{k}_z^2)^{1/2}$, $\tilde{k}_i \equiv k_i/k_f$, and $\tilde{\omega} \equiv \omega/(k_f U)$.
In the specific case of our study, where $\BB = B_0 \bv{\hat{x}}$ and $\OO = \Omega \bv{\hat{z}}$, this simplifies to:
\begin{equation}
    \tilde{\omega}\left(\bv{\tilde{k}}; \Lambda\right) = -\frac{1}{2}\frac{\tilde{k}_z}{\Lambda \tilde{k} Ro} \pm \frac{1}{2}\sqrt{\left(\frac{\tilde{k}_z}{\tilde{k} Ro}\right)^2 + 4 \left(\frac{\tilde{k}_x}{M}\right)^2}. \label{eq:disprel2}
\end{equation}
See Figure \ref{fig:waves} for a visualization of this dispersion relation, which depends on $\tilde{k}_x, \tilde{k}_z,$ and $\tilde{k}$.


\bibliographystyle{jfm}
\bibliography{main}

\end{document}